\documentclass{article}

\usepackage[affil-it]{authblk}

\usepackage{graphicx}
\usepackage{amsmath}
\usepackage{amsxtra}
\usepackage{amstext}
\usepackage{amssymb}
\usepackage{latexsym}
\usepackage{multirow}
\usepackage{lscape}
\usepackage{epstopdf}
\usepackage{setspace}
\usepackage{calc}
\usepackage{color}
\usepackage{multicol}
\usepackage{pdfpages}
\usepackage{etex}
\usepackage{morefloats}
\newcommand{\dpart}[2]{\frac{\partial #1}{\partial #2}}
\newcommand{\ud}{\mathrm{d}}
\newcommand{\vect}[1]{\pmb{#1}}
\newcommand{\CO}{CO$_2$\:}

\begin{document}

\title{A New Equation of State for CCS Pipeline Transport: Calibration
  of Mixing Rules for Binary Mixtures of CO$_2$ with N$_2,$ O$_2$ and H$_2$}
\author{Thomas A. Demetriades}
\affil{KBC Advanced Technologies, Westmead House, Farnborough, Hampshire, GU14 7LP.}
%\ead{tdemetriades@kbcat.com}

\author{Richard S. Graham}
\affil{School of Mathematical Sciences, University of Nottingham, Nottingham NG7 2RD, UK.}

\maketitle

\begin{abstract}
 One of the aspects currently holding back commercial scale deployment
 of carbon capture and storage (CCS) is an accurate understanding of the
 thermodynamic behaviour of carbon dioxide and relevant impurities
 during the pipeline transport stage.  In this article we develop a general framework for deriving
pressure-explicit EoS for impure CO$_2$. This flexible framework facilitates
ongoing development of custom EoS in response to new data and computational
applications. We use our method to generalise a recent EoS for pure
\CO [Demetriades {\em et al.} Proc IMechE Part E, 227 (2013) pp. 117] to binary
mixtures with N$_2$, O$_2$ and H$_2$, obtaining model parameters by fitting
to experiments made under conditions relevant to CCS-pipeline
transport. Our model pertains to pressures up to 16MPa and
temperatures between 273K and the critical temperature of pure
CO$_2$.   In this region, we achieve close agreement with experimental
data. When compared to the GERG EoS, our EoS has
a comparable level of agreement with \CO-N$_2$ VLE experiments
and demonstrably superior agreement with the O$_2$ and H$_2$ VLE data.
Finally, we discuss future options to improve the calibration of
EoS and to deal with the sparsity of data for some impurities.
\end{abstract}

%\end{frontmatter}

\section{Introduction and Background}

\subsection{Carbon capture and storage}
Carbon capture and storage is a crucial technology in the
international efforts to meet carbon dioxide emission targets \cite{BootHandford:2014p4173}. Capturing \CO from industrial sources can lead to a $90\%$ reduction in emissions. However, no gas separation process is $100\%$ efficient, and as a result the \CO generated from power generation or by industry can contain a number of different impurities, depending on its source. These impurities can, depending on their composition and concentration, greatly influence the physical properties of the fluid compared to pure CO$_2$. Impurities have important design, safety and cost implications for the compression and transport of \CO and its storage location, for example geological sequestration.
This research is designed to tackle one of the key technical
challenges facing the development of commercially viable \CO transport
networks: modelling physical behaviour of impure CO$_2$, under
the conditions typically found in carbon capture from power stations,
and in high-pressure (liquid phase) and low-pressure (gas phase)
pipelines. Accurate modelling of the physical properties of \CO
mixtures is essential for the design and operation of compression and
transport systems for CO$_2$.
In particular the variation of fluid
density and phase-behaviour with temperature and pressure is key to
many CCS processes. For example, pipeline transport of \CO is only
viable if the fluid remains in the homogenous phase. 
%\CO fluid
%behaviour  is greatly influenced by the many impurities that result from power generation.

There has been recent work to define the expected operating conditions
for CCS pipelines \cite{Seevam07,deVisser08}. The most efficient way of transporting \CO is in the homogenous phase,
at pressures in the vicinity of its critical point.
For the transport temperature, the upper temperature will be set by
the compressor discharge temperature and the temperature limits of the
pipeline coating and the lower temperature will correspond to the
winter ground temperature of the surrounding soil\cite{Mohitpour2012}. Expected impurity levels are
about $\lesssim 4\%$, with  N$_2$ , O$_2$ and H$_2$ being key impurities \cite{Seevam07,deVisser08,Porter2015,Tenorio2015}.  This
range of pressure, temperature and impurity level
define pipeline operating conditions and provide a target window for
CCS-oriented modelling. However, CCS-relevant models should aspire to model a
wider range of conditions, particularly for the impurity level, for
the following reasons. Coexistence leads to the formation of a vapour phase that is
considerably richer in impurity than the overal mixture. Upset
conditions, in which a greater concentration of impurity is accidentally
introduced
into the \CO stream, must be understood and mitigated for. Finally, an effective way to ensure physical robustness of the model is to test for a
wider range of impurity conditions.

\subsection{\CO modelling in CCS}

In models of the CCS process, the fluid behaviour of \CO~mixtures is typically predicted by an equation of state (EoS). EoS
vary in their mathematical form, accuracy, region of validity and
computational complexity. Because different applications have
different requirements there is no single EoS that is ideal for all
applications. In particular, there is a balance between mathematical
simplicity and accuracy of prediction. To optimise their accuracy, EoS need to be calibrated by
fitting their parameters to experimental measurements on \CO mixtures. However, new measurements become available very
frequently, offering the opportunity to improve the models. Thus,
there is an ongoing need to regularly rederive, refine and
reparameterise EoS. Currently, the inability to rapidly assimilate
ongoing measurements into suitable EoS delays or prevents knowledge gained from
experiments from being applied in CCS modelling.

A pressure-explicit EoS is an expression for a fluid's  pressure
$P^*$, as a function of molar volume $v^*$ and temperature $T^*$. Usually, the terms in the model
that describe the deviation from ideal gas behaviour are empirically
postulated. A widely used example  is the
Peng Robinson equation \cite{Peng76}. Such EoS gives an
explicit prediction of the pressure-volume curve for homogeneous
fluids and can also predict the coexistence behaviour. For
thermodynamic co-existence,  the coexisting phases must
have matched
fugacity for each chemical species, which is equivalent to matching the chemical potential. This 
 translates into a constraint involving the integral of $P^*$ over volume, at constant temperature. 
Thus, the coexistence behaviour can be predicted by numerically
searching for two volumes that obey both the EoS and thermodynamic
coexistence requirement. Typically, EoS contain empirical parameters that are estimated  by fitting the EoS to experimental data for the pure material. These parameter are generalised to mixtures through a set of empirical mixing rules that extend the EoS to multi-component mixtures. These mixing rules also allow calculation of the phase behaviour, through a generalised expression for the fugacity. The pure component parameters and the mixing rules may vary with temperature.

There is an alternative method to formulate EoS, in which the
volume and temperature-dependence of the Helmholtz
free energy, rather than the pressure, is postulated\cite{Span96,Gerg2004}. Similarly to pressure-explicit EoS, these
models contain empirical terms that described the deviation from ideal
gas behaviour. The two formulations are, however,
mathematically equivalent as integration of a pressure-explicit EoS
leads to an expression for the Helmholtz free energy. In this work, we use
the pressure-explicit formulation, as measurements of pressure can be
directly visualised and so postulating empirical terms for $P^*(v^*)$ is
physically more intuitive than for the free energy.

%The typical current approach to EoS modelling is to specify a particular 
%form for the function $ P = P(v)$; a form for its parameters 
%$(a,b,...)$ as functions of T; and a mixing rule for each 
%parameter.  Then the model is calibrated by minimising, with respect to 
% the parameters, some measure of the deviation between model and 
%data.  Predictions are then made from the calibrated model by `plugging-in' the estimated parameter values. This approach has a number of significant limitations, which prevents the practical, effective exploitation of EoS in CCS modelling.

%\subsection{EoS for carbon dioxide}

There is considerable uncertainly over which EoS is most appropriate
for CCS modelling. Although the Peng-Robinson model is mathematically simple and
numerically cheap, it was derived to compute separation of mixtures
for the natural gas industry, where CO$_2$ is a minor additive. Thus,
its parameters are optimised to the coexistence behaviour,
particularly for the gas phase. Agreement with density measurements
for pure CO$_2$ around the critical pressure, which is key to CCS
modelling, is unacceptably poor. Moreover,
the mixing rules are not optimised to CCS-relevant mixtures. Variants
of the Peng-Robinson model usually suffer the same limitations.

For pure CO$_2$, the Span-Wagner EoS\cite{Span96} covers from the triple-point
temperature up to very high pressures and temperatures with very high
accuracy. Furthermore an EoS by Yokozeki\cite{Yokozeki:2004p4207}
captures solid-liquid coexistence of pure CO$_2$. Also for pure CO$_2$, there
is a recent composite EoS\cite{Wareing:2013p4029} that combines the Peng-Robinson model for the
gas phase and accurate tabulated data for the solid and liquid
phases\cite{Span96,MartinTrusler:2011p4209,Jager:2012p4208}. There
have also been studies comparing different pure \CO EoS for CCS
applications \cite{Wareing:2013p4029,Han:2008p4215}.
There are complex and accurate EoS for \CO mixtures, including the
SAFT\cite{Chapman89}, PC-SAFT\cite{Gross:2001p4212},
GERG\cite{Gerg2004} and EOS-CG\cite{Span:2013p4203,GenertThesis2013}
models. The SAFT, PC-SAFT and EOS-CG models have recently been compared to
some CCS-relevant
measurements\cite{Diamantonis:2013p4206,GenertThesis2013}. However,
they have not been compared to recently emerged VLE data for \CO-H$_2$ mixtures\cite{Fandino:2015p4210,Tenorio2015}.
Furthermore, the mathematical complexities of these models preclude their use in some CCS applications.

There is clearly considerable scope to derive new EoS, specifically
for CO$_2$ mixtures, retaining the simplicity of Peng-Robinson type
equations but with improved quantitative performance for CCS
applications. Indeed, a simplified version of the Span-Wagner model
for pure \CO has been produced\cite{Span:2003p4214}. However, the measurements against
which a CCS-focussed EoS should be calibrated continually evolve as CCS-oriented projects deliver new data to complement the literature data. Additionally, every user's requirements differ, depending on  the accuracy demanded, the region of interest, and the computational complexity that can be tolerated.
Thus, there is no single ``silver-bullet'' EoS, suitable for all CCS
applications and all users. Instead, each application's requirements
can only be met by a bespoke EoS. There is a need for a flexible,
general framework to derive and parameterise new EoS in response to
emerging measurements and computational applications.

\subsection{Aims of this work}

In this work we introduce a general and flexible framework for
deriving new pressure-explicit EoS. We derive an expression for the
mixture fugacity for an arbitrary EoS with arbitrary mixing rules.
This allows an EoS to be generalised to mixtures, for any choice of
mixing rules, without needing to recompute the fugacity integral. Therefore,
our approach allows convenient
modification of the form of EoS and mixing rules, in response to new
measurement or new physical insights. We demonstrate the flexibility
and effectiveness our approach by generalising a recent
pressure-explicit EoS for pure \CO to mixtures with N$_2$, O$_2$ and
H$_2$ and then calibrate this
model against data for CCS-relevant mixtures, focussing on
pressure-volume behaviour and coexistence. Our approach is readily
generalisable to other common CCS impurities, such as argon and methane
and can also fit to other thermodynamic variables such as enthalpy and
heat capacity.

\section{A general expression for the mixture fugacity in pressure-explicit equations of state}
In this section we consider a general pressure-explicit EoS and derive
results for arbitrary EoS and mixing rules. To
compute coexistence we require an expression for the fluid
fugacity. For pure fluids this can be obtained by directly integrating
the EoS, using eqn~\eqref{eq:3a}. Whether this integral can be
performed explicitly depends on the expression chosen for the
EoS. Furthermore, for the mixture fugacity, one must compute a more complicated
integral, involving the mixing rules (eqn~\eqref{eq:1}). Computing this
mixture integral directly is not mathematically
convenient for all but the simplest EoS.
In this section we show that, provided the pure fluid integral can be
performed, then a closed formed expression for the mixture fugacity
integral can always be found. Furthermore, we derive a general
expression for the mixture fugacity in terms of derivatives of the
mixing rule.

We begin with an EoS in the form
$P^\star=P^\star(v^\star,\vect{\theta}^\star,T^\star)$, where
$P^\star$ is the pressure in Pa, $v^\star$ is the molar volume in
mol/m$^3$, $T^\star$ is temperature in K and $\vect{\theta}^\star$ is
a vector of model parameters. Here, stars denote dimensional
quantities. The fugacity co-efficient for the pure fluid, $\ln\phi$,
is given by eqn~\eqref{eq:3a} in \ref{sec:mixt-fugac-equat}. The EoS
is generalised to mixtures by allowing the parameter $\vect{\theta}$
to depend on the fluid composition, $\vect{x}$, to be specified via a mixing rule. Here $x_i$ denotes the mol fraction of species $i$ of the mixture. Leaving the functional forms of $P^\star$ and $\vect{\theta}^\star$ unspecified, we show in \ref{sec:mixt-fugac-equat} that the fugacity coefficient of species $i$ in a mixture is given by
\begin{equation}
  \label{eq:14} \ln\bar{\phi}_i(v^*,\vect{x})=\ln\phi+\sum_{j=1}^{Np}\dpart{F}{\theta^\star_j}\left(\dpart{\theta^\star_j}{x_i}-\sum_{k=1}^{N_{sp}}x_k\dpart{\theta^\star_j}{x_k}\right)
,
\end{equation}
where $F=\ln\phi-((Z-1)-\ln Z)$. We see that, if the fugacity integral
can be performed for the pure fluid, then eqn~\eqref{eq:14} ensures
that the fugacity integral for mixtures can also be written as a
closed form expression. Finally, we note that eqn~\eqref{eq:14} is
invariant under a constant scaling of any of the model parameters. This means that the expression is
completely unchanged by any non-dimensionalisation that merely scales
the model parameters by a constant.

\section{A new Equation of State for impure \CO}
\subsection{Pure CO$_2$}
We begin with a recently proposed EoS for pure \CO by Demetriades {\em
  et al.}\cite{Demetriades13}. The dimensional EoS is
\begin{equation}\label{equation:DimEoS}
P(v^\star, T^\star; \vect{\theta}^\star) = \frac{RT^\star}{v^\star + a^\star} - \frac{b^{\star 2}}{v^{\star 2} + c^{\star 2}} - \frac{d^{\star 3}}{v^{\star 3} + e^{\star 3}} + \bigg(\frac{f^\star}{v^\star - g^\star}\bigg)^6,
\end{equation}
where $R$ is the ideal gas contant and $\vect{\theta}^\star=(a^\star,...,g^\star)$ is the vector of model parameters. We perform the following non-dimensionalisation, for the physical variables
\begin{eqnarray}\label{equation:NonDimensionalisation}
P^\star = P_c^\star P,\;\;\;T^\star &=& T_c^\star T, \;\;\; v^\star = \bigg(\frac{R T_c}{P_c}\bigg) v, 
\end{eqnarray}
and for the model parameters
\begin{eqnarray}\label{equation:NonDimensionalisation2}
a^\star = \bigg(\frac{R T^\star_c}{P^\star_c}\bigg) a, \;\;
b^\star = \bigg(\frac{(R T^\star_c)^2}{P^\star_c}\bigg) b, \;\;
c^\star = \bigg(\frac{R T^\star_c}{P^\star_c}\bigg)^2 c, \;\;
d^\star = \bigg(\frac{(R T^\star_c)^3}{P^{\star 2}_c}\bigg) d,\nonumber\\
e^\star = \bigg(\frac{R T^\star_c}{P^\star_c}\bigg)^3 e, \;\;
f^\star = \bigg(\frac{R T^\star_c}{P^{\star 5/6}_c}\bigg) f, \;\;
g^\star = \bigg(\frac{R T^\star_c}{P^\star_c}\bigg) g. 
\end{eqnarray}
Here $P^\star_c$ and $T_c^\star$ are the critical pressure and temperature of pure CO$_2$, respectively. After non-dimensionalisation our EoS becomes
\begin{equation}\label{equation:NonDimEoS}
P(v, T) = \frac{T}{v + a} - \frac{b^2}{v^2 + c^2} - \frac{d^3}{v^3 + e^3} + \bigg(\frac{f}{v - g}\bigg)^6.
\end{equation}
and the fugacity integral for pure CO$_2$ (eqn~\eqref{eq:3a}) becomes
\begin{equation}\label{equation:NonDimFugPureConstraint}
\ln \phi = \int_\infty^v \bigg( \frac{1}{v'} - \frac{ P(v',T)}{T} \bigg) \mathrm{d}v' - \ln \bigg( \frac{Pv}{T} \bigg) + \bigg( \frac{Pv}{T} \bigg) - 1,
\end{equation}
which can be evaluated by substituting in eqn~\eqref{equation:NonDimEoS}
\begin{eqnarray}
\ln \phi(v) &=& \ln \bigg(\frac{v}{a + v} \bigg) + \frac{b^2}{T c} \arctan \bigg( \frac{v}{c} \bigg)
\nonumber \\
& & + \frac{d^3}{3T e^{2}} \ln \bigg( \frac{e+v}{\sqrt{e^{2}-e v + v^2}} \bigg) + \frac{d^3}{T \sqrt{3} e^2} \arctan \bigg( \frac{2 v - e}{\sqrt{3} e} \bigg) - \frac{f^6}{5T (g-v)^5}\nonumber \\
& & - \frac{\pi}{2T} \bigg( \frac{b^2}{c} + \frac{d^3}{\sqrt{3} e^{2}} \bigg) - \ln \bigg( \frac{P(v) v}{T} \bigg) + \bigg( \frac{P(v) v}{T} \bigg) - 1.
\end{eqnarray}

We repeated the fitting procedure for pure CO$_2$ from Demetriades
{\em et al.}\cite{Demetriades13} to obtain slightly improved agreement
with the pure CO$_2$ data over our previous article. The resulting variation of model parameters with temperature is given by
\begin{eqnarray}
a_{\mathrm{CO_2}}(T) &= &
| T-1 | ^{0.626207} (33.9261 | T-1 | ^2 - 8.10461 | T-1 | 
+ 0.805812 ) + 0.2712941,%1630837575312 
\nonumber\\
b_{\mathrm{CO_2}}(T)& = &
| T-1 | ^{0.405254} ( - 13.5708 | T-1 | ^2 + 4.48534 | T-1 | - 0.295229 )
+ 0.3326169,%0686926754767
\nonumber\\
c_{\mathrm{CO_2}}(T) &=& 
| T-1 | ^{0.515789} ( - 3.77054 | T-1 | ^2 + 1.72673 | T-1 | 
{} - 0.478733 ) + 0.238762,%25996324883019 
\nonumber\\
d_{\mathrm{CO_2}}(T) &=& 
| T-1 | ^{1.27068} ( 0.000634507 | T-1 | ^3 - \text{8.327888244017052E-6} | T-1 | ^2 \nonumber \\
&& - 0.0000382867 | T-1 | + \text{4.661593764290955E-6} ) {} 
{} - 0.000374407355,%02740741443 
\nonumber\\
e_{\mathrm{CO_2}}(T)& =& 0.780746514,%36175671059,
\label{eq:4}\\
f_{\mathrm{CO_2}}(T)& =& 
| T-1 | ^{0.192269} ( 0.210429 | T-1 | ^2 - 0.199813 | T-1 | 
{} + 0.0528131 ) + 0.0787701,%14175749938219 
\nonumber\\
g_{\mathrm{CO_2}}(T) &=& 
| T-1 | ^{0.198411} ( - 0.185594 | T-1 | ^2 + 0.0931741 | T-1 |
- 0.0510056 ) + 0.074028115.%340316985540 
\nonumber
\end{eqnarray}
Thus each parameter was fully defined and this completed the model in
the case of pure CO$_2$. We note that the parameter variations
given here differ slightly from those proposed in our published work
\cite{Demetriades13}. The model proposed here has a better accuracy
than that of our prior version. We now fix these model parameters for the remainder of
this article.

\subsection{Generalisation to impure CO$_2$}
We generalise our model to impure \CO by allowing the model parameters
to depend on the impurity concentration, $x_{\mathrm{imp}}$.
We opted to impose a linear mixing rule for each of the model parameters in equation \eqref{equation:NonDimEoS}. For parameter $a$, the mixing rule for a binary mixture of CO$_2$ and a single impurity (imp) is:\begin{equation}\label{equation:CO2N2MixingRules}
a(T) = x_{\mathrm{CO_2}} a_{\mathrm{CO_2}}(T) + x_{\mathrm{imp}} a_{\mathrm{imp}}(T),
\end{equation}
where $a_{\mathrm{CO_2}}(T)$is defined in equation~\eqref{eq:4} for
pure \CO and $a_{\mathrm{imp}}(T)$ is the analogous quantity for the impurity, which we need to specify. Replacing $a$ in eqn~\eqref{equation:CO2N2MixingRules} with any of the other model parameters provides the mixing rule for that parameter.  We note at this stage that for a binary system
\begin{equation}
x_{\mathrm{imp}} = 1 - x_{\mathrm{CO_2}}.
\end{equation}
From here on we use $x$ to denote the impurity fraction.

\subsection{Fitting strategy for impurity data}
Our fitting strategy for the impurity model is similar to our
methodology for pure \CO \cite{Demetriades13}, in that we fit model
parameters at different temperatures by using simulated annealing, 
which searches for global minima and has the ability to escape local
minima. 
%as implemented in the Mathematica software. 
The correct coexistence behaviour is
ensured by imposing a term in the minimisation that penalises
differences in the predicted fugacity at the experimentally determined
coexistence points (see eqn~\eqref{eq:6}). This approach requires
vapour-liquid equilibrium (VLE) measurements, for both the coexisting
mol fractions and molar volumes at each temperature to be fitted.

\section{Carbon Dioxide--Nitrogen Binary Mixture\label{section:Fitting the CO$_2$--N$_2$ Model Parameters}}
\subsection{Data Availability}

\begin{table}[htb]
\begin{spacing}{1.0}
\begin{center}
 \begin{tabular}{| l | l | l |}
\hline \hline
Temperature (K) & Homogeneous Density & VLE Data \\
\hline \hline
260		& \cite{Bailey89}, \cite{Ely87}, \cite{Ely89} & \\
265		& \cite{Brugge97}, \cite{Duarte-Garza95-II} & \\
270		& \cite{Ely87}, \cite{Ely89} & \cite{Brown89-I}$^{ \dagger }$, \cite{Brown89-II}$^{ \dagger }$, \cite{Somait78}$^{ \dagger }$, \cite{Yucelen99}$^{ \dagger }$ \\
273.15	& {\bf \cite{Arai71}}, \cite{Bailey89}, \cite{Kritschewsky40} & \cite{Arai71}, \cite{Kaminishi66}$^{ \dagger }$, {\bf \cite{Muirbrook64,Muirbrook65}},\cite{Yorizane85}$^{ \dagger }$ ,\\ & & \cite{Tsiklis46}$^{ \dagger }$, \cite{Weber84}$^{ \dagger }$, \cite{Yorizane70}$^{ \dagger }$, \cite{Zenner63}$^{ \dagger }$  \\
275		& \cite{Brugge97}, \cite{Duarte-Garza95-II}, \cite{Ely87}, \cite{Ely89}, \cite{Mondejar12} & \\
280		& \cite{Ely87}, \cite{Ely89} & \\
285		& \cite{Brugge97}, \cite{Duarte-Garza95-I}, \cite{Duarte-Garza95-II}, \cite{Ely87}, \cite{Ely89} & \\
288.15	& {\bf \cite{Arai71}} & {\bf \cite{Arai71}}, \cite{Kaminishi66}$^{ \dagger }$, \cite{Xu92}$^{ \dagger }$ \\
288.706	& \cite{Bailey89} & \\
290		& \cite{Ely87}, \cite{Ely89} & \\
293.3	& \cite{Jiang90} & \cite{Xu92}$^{ \dagger }$ \\
295		& \cite{Ely87}, \cite{Ely89} & \\
298.15	& \cite{Haney44} & \cite{Kaminishi66}$^{ \dagger }$  \cite{Yorizane85}$^{ \dagger }$  \\
300		& \cite{Bailey89}, \cite{Brugge89}, \cite{Brugge97}, \cite{Duarte-Garza95-I}, \cite{Duarte-Garza95-II}& \\
&\cite{Ely87}, \cite{Ely89}, \cite{Esper89}, \cite{Mondejar12} &\\
301.3	& & \cite{Bian93} \\
303.3	&\cite{Mantovani12}  & \cite{Bian93} \\
\hline
\multicolumn{3}{| c |}{$^{ \dagger }$ means the VLE data in this source do not contain volumes} \\
\hline \hline
\end{tabular}
\caption{A summary of available data for \CO--N$_2$
  mixtures. Experimental data used in
  our parameter fitting are in boldface. \label{table:CO2N2DataSummary}}
\end{center}
\end{spacing}
\end{table}

 We begin this section by noting the literature data available to fit
 the CO$_2$--N$_2$ mixture, which is summarised in Table
 \ref{table:CO2N2DataSummary}. As specified in above, we require VLE
 data, including coexisting volumes, ideally with homogeneous phase
 density data all at the same temperature.
There are only two relevant temperatures in the literature where this full
spread of data exist and, in some cases, the fraction of N$_2$ used
for the homogenous phase measurements is outside the range of
relevance to CCS.
 In order to find the pure N$_2$ parameter values, we fitted our model
 to the \CO-N$_2$ mixture data at 273.15K and 288.15K, choosing these
 temperature as they both had measurements of the coexisting volumes
 and homogenous phase. A summary of the data used for fitting is as
follows. For 273.15K we used VLE data by Muirbrook {\em et al.}\cite{Muirbrook64,Muirbrook65} and
homogenous phase data by Arai {\em et al.}\cite{Arai71} for $x_{N_2}$
in the range $7.1-25.4\%$. For 288.15K we used VLE data by Arai {\em et al.}\cite{Arai71}  and
homogenous phase data by the same authors for $x_{N_2}$
in the range $6.0-10.3\%$.

\subsection{Fitting the \CO--N$_2$ Binary Model Parameters}
 We proceeded first by determining the pure N$_2$ parameters
 $a_{\mathrm{N_2}}$ to $g_{\mathrm{N_2}}$ at the available temperature
 points by a similar series of optimisations as used in the pure \CO
 case. Specifically, we minimised an error function, which in the case
 of a binary mixture, was  comprised of two contributions: from the
 homogeneous measurements 
 \begin{equation}
   \label{eq:4a}
   E_{\mathrm{Hom}}=\sum_{i=1}^{N_h} \bigg( \frac{P(v_{\mathrm{DATA}}^{(i)}, T, x_{\mathrm{DATA}}^{(i)}) - P_{\mathrm{DATA}}^{(i)}}{P_{\mathrm{DATA}}^{(i)}} \bigg)^2, 
 \end{equation}
where $N_h$ is the number of homogeneous measurements at this temperature, $v_{\mathrm{DATA}}^{(i)}$ and $x_{\mathrm{DATA}}^{(i)}$ denote the volume and $N_2$ composition, respectively, of the $i$th homogenous data point and $P(v,T,x)$ is the model's predicted pressure. The second contribution comes from the VLE measurements
\begin{eqnarray}
  \label{eq:6}
  E_{\mathrm{VLE}}&=&\sum_{j=1}^{N_v} \bigg[ 
  ( P(v_{\mathrm{vap}}^{(j)}, T, x^{(j)}_{\mathrm{vap}}) - P_{\mathrm{vap}}^{(j)} )^2 {} 
+ ( P(v_{\mathrm{liq}}^{(j)}, T, x^{(j)}_{\mathrm{liq}}) - P_{\mathrm{vap}}^{(j)} )^2 \nonumber \\ 
& &+ ( \ln \bar{\phi}_{\mathrm{CO_2}}(v_{\mathrm{vap}}^{(j)},x^{(j)}_{\mathrm{vap}}) - \ln \bar{\phi}_{\mathrm{CO_2}}(v_{\mathrm{liq}}^{(j)},x^{(j)}_{\mathrm{liq}})+\ln((1-x^{(j)}_{\mathrm{vap}})/(1-x^{(j)}_{\mathrm{liq}})) )^2 {} \nonumber\\ 
&&+ ( \ln \bar{\phi}_{\mathrm{N_2}}(v_{\mathrm{vap}}^{(j)},x^{(j)}_{vap}) - \ln \bar{\phi}_{\mathrm{N_2}}(v_{\mathrm{liq}}^{(j)},x^{(j)}_{liq}) +\ln(x^{(j)}_{\mathrm{vap}}/x^{(j)}_{\mathrm{liq}}) )^2 {} \bigg],
\end{eqnarray}
where $N_v$ is the number of VLE measurements at this temperature,
$v_{\mathrm{vap/liq}}^{(j)}$ and $x_{\mathrm{vap/liq}}^{(j)}$ denote the coexisting liquid/vapour volume and $N_2$ composition, respectively, of the $j$th VLE measurement, $P_{\mathrm{vap}}^{(j)}$ is the $j$th vapour pressure measurement and  $\ln \bar{\phi}$ is the model's predicted fugacity. The two contributions were combined into a total error function
\begin{eqnarray}\label{equation:ErrFnMixCO2N2}
E_{\mathrm{Tota}l} = W E_{\mathrm{Hom}}+(1-W)E_{\mathrm{VLE}},
\end{eqnarray}
where $W$ is a parameter that determines the weighting of the fitting
between homogenous and VLE data. We chose  $W=0.1$, to balance fitting between the VLE and
 homogenous phase data. This generated values for each of the seven pure N$_2$
 parameters at both of the temperature points.

\subsection{Variation of Model Parameters with Temperature}
\begin{table}
\begin{center}
 \begin{tabular}{| c || c | c  |}
\hline 
$\theta$ &  $\alpha_0$& $\alpha_1$ \\
\hline
$a_{\mathrm{N_2}}$ &-2.94804 &3.14877 \\
   $b_{\mathrm{N_2}}$ &0.616547 &-0.654773 \\
$c_{\mathrm{N_2}}$ &-4.83602 &5.83048 \\
$d_{\mathrm{N_2}}$ & 4.69708  &-5.23021 \\
$e_{\mathrm{N_2}}$ & 6.23314 & -6.56762\\
$f_{\mathrm{N_2}}$ &-11.6889 &13.4184 \\
$g_{\mathrm{N_2}}$ &15.3391 &-17.2528 \\
\hline
\end{tabular}
\caption{Coefficients describing the temperature dependence of the
  N$_2$ model parameters in equation~\eqref{equation:LinearFit}. \label{table:TempDep}}
\end{center}
\end{table}

We had determined values for each parameter at each of the two temperatures. To obtain an expression for the temperature dependence of each parameter we converted our parameter values into a linear dependence on temperature,
\begin{equation}\label{equation:LinearFit}
\theta_{\mathrm{N_2}}(T) = \alpha_0+ \alpha_1 T,
\end{equation}
where $\theta$ is a model parameter. The values of $\alpha_0$ and
$\alpha_1$ we obtained for each of the seven N$_2$ parameters  are in table~\ref{table:TempDep}

\subsection{Performance of the Model for Binary Mixtures of \CO and N$_2$}

\begin{figure}[htb]
\begin{center}
\includegraphics[scale=0.4]{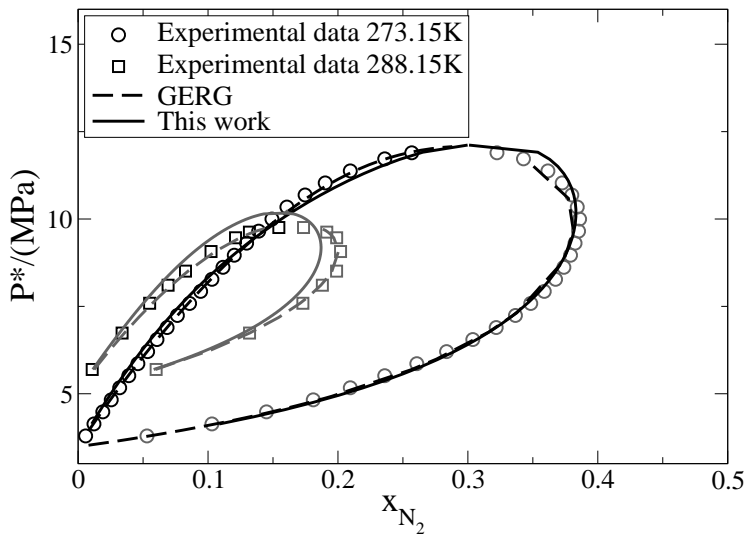}
\caption{Comparison of experiments and our model for the VLE
  coexisting mole fraction for \CO-N$_2$ mixtures. Experimental data at 273.15K from
  Muirbrook {\em et al.}\cite{Muirbrook64,Muirbrook65} and at 288.15K from Arai {\em
    et al.}\cite{Arai71}. Also shown are the results from the GERG EoS
\cite{Gerg2004}.
%; at 293K from Xu {\em et al.}\cite{Xu92}; and at 301.3K from Bian {\em
%    et al.}\cite{Bian93}.
\label{figure:CO2N2CoexxP}}
\end{center}
\end{figure}

\begin{figure}[htb]
\begin{center}
\includegraphics[scale=0.4]{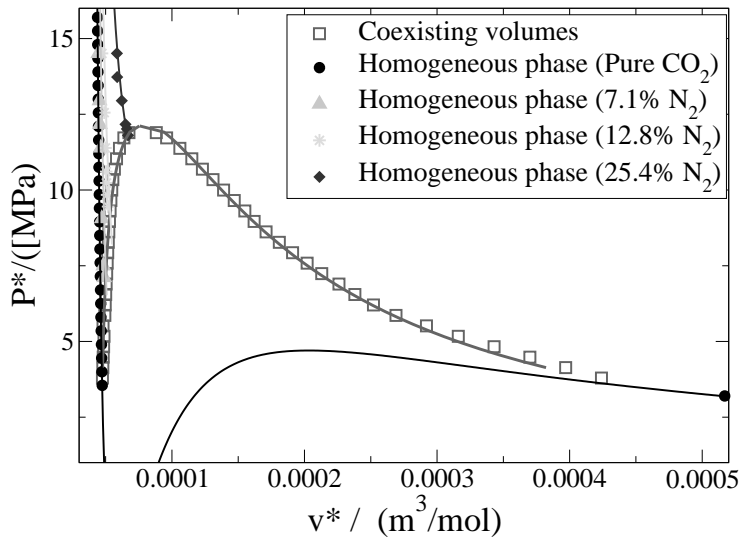}
\caption{Comparison of experiments and our model for the
  pressure-volume behaviour of \CO-N$_2$ mixtures at 273.15K in both
  coexistence and the homogenous phase. Experimental data from
  Muirbrook {\em et al.}\cite{Muirbrook64,Muirbrook65} (coexistence)
  and from Arai {\em
    et al.}\cite{Arai71} (homogenous phase). \label{figure:CO2N2CoexvP}}
\end{center}
\end{figure}

\begin{figure}[htb]
\begin{center}
\includegraphics[scale=0.4]{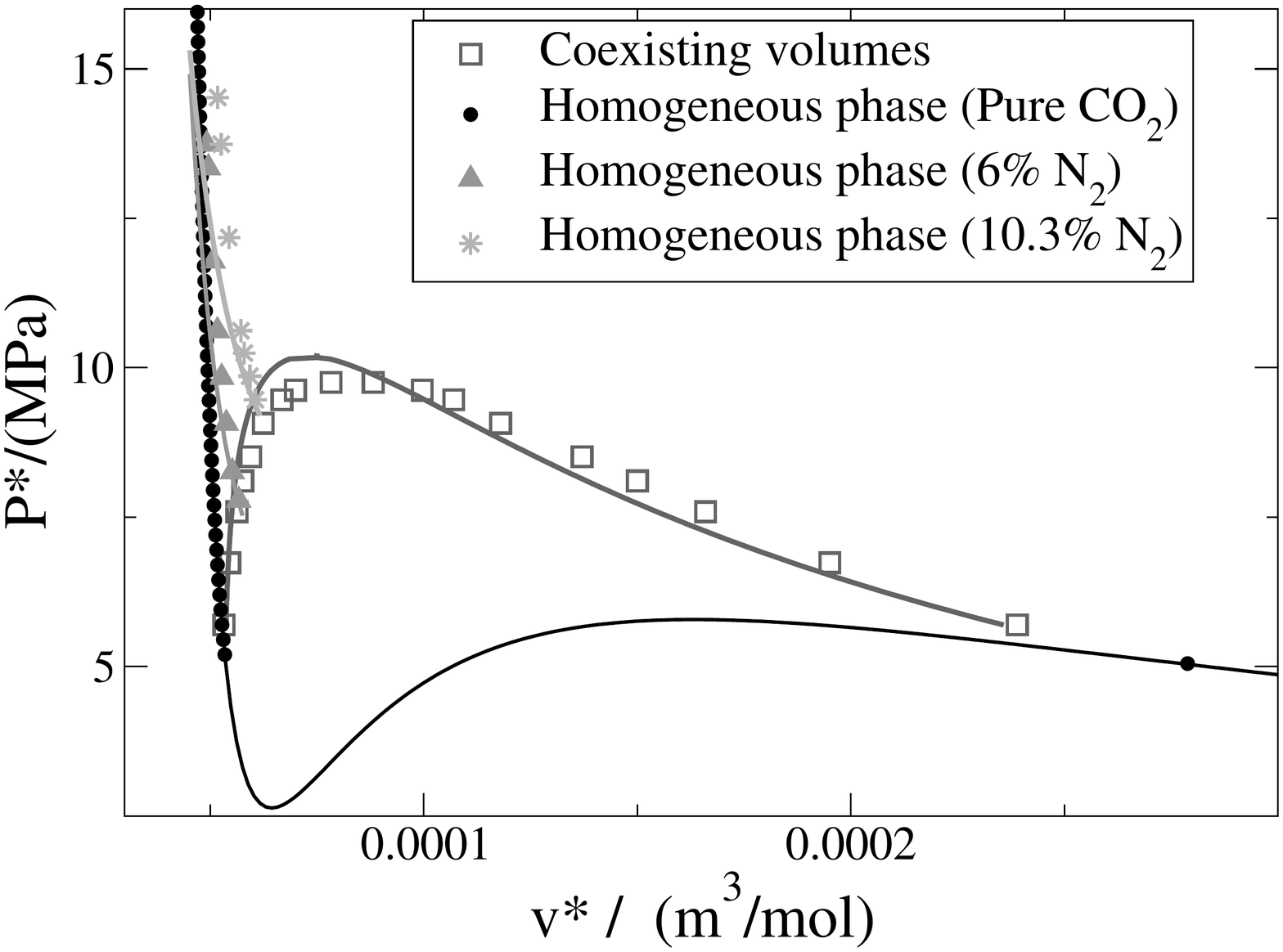}
\caption{Comparison of experiments and model predictions for the
  pressure-volume behaviour of \CO-N$_2$ mixtures at 288.15K in both
  coexistence and the homogenous phase. Experimental data from Arai {\em
    et al.}\cite{Arai71}. \label{figure:CO2N2CoexvP_288K}}
\end{center}
\end{figure}

 With the model thus fully defined for the binary system involving
\CO and N$_2$, we were able to assess its
 performance. We did this by comparing the
 completed model to the data was used to generate the
 fitted parameters. 
%We also tested the model predictions at 301.3K,
 %against data that had not been used to fit the model.
We computed the model's coexistence predictions, at a given
temperature, as follows. We began at the pure \CO coexistence
pressure, where the coexisting volumes are known and the coexisting
mol fractions are both $100\%$ CO$_2$. We then increased the pressure by
a small increment and searched for the mol fraction and volume of the
liquid and vapour that satisfy the following four coexistence
conditions: the liquid phase pressure equals the prescribed pressure; the
vapour phase pressure matches the prescribed pressure; the \CO  fugacity of the liquid matches that of the vapour; and the N$_2$
fugacity of the liquid matches that of the vapour. We used a
numerical non-linear root-finding algorithm to locate the co-existence
point. Once the coexistence root was found we increased slightly the
pressure and repeated the root finding, using the previous root as an
initial guess for the new root. We repeated this process, moving up in
small pressure increments, until the
critical point was reached.

A comparison between the experiments and our model for the coexisting
mol fraction is
shown in figure~\ref{figure:CO2N2CoexxP}. Also shown are the results from the GERG
EoS. At both temperatures for both models the agreement is very
good. Our model performs similarly to the GERG, except at pressures approaching the critical
point for 288.15K. Here, our model fails slightly to capture accurately the approach
to the critical point, which is most clearly seen by the rapid
 convergence of the liquid and vapour mol fraction as the experiments
 approach the critical pressure, a feature that the GERG captures.
For coexisting volumes (
figures~\ref{figure:CO2N2CoexvP} and \ref{figure:CO2N2CoexvP_288K})
our model performs very well, except for close to the critical point
at 288.15K,
where our model slightly over-predicts the pressure.
Our model captures the mixtures' pressure-volume data for the
homogenous phase (figures~\ref{figure:CO2N2CoexvP}), except at
high pressures  at 288.15K where the model's results are slightly
closer to the pure \CO behaviour than the experiments.
Finally, we also highlight that the best performance of our model
 occurred for the liquid coexistence properties. This will ensure that the
 model accurately predicts the liquid saturation line, which is of
 particular relevance to CCS
 pipelines as it is the edge of the homogeneous phase region and,
 ultimately, it defines the minimum safe pipeline operating pressure
 for a given temperature and overall impurity composition.

Overall,  this mixture shows the promise of our
methodology. In the areas where our current model fails a little to fully
capture the data, namely close to the critical pressure, the
flexibility and generality of our approach will enable future work to
readily introduce refined mixing rules or terms in the original EoS to
obtain further improved agreement. Conversely, our framework will also
enable the complexity of the EoS to be reduced for applications that
demand a simple EoS but can compromise on the closeness to experiments.

\section{Carbon Dioxide--Oxygen Binary Mixture\label{section:Fitting the CO$_2$--O$_2$ Model Parameters}}
\subsection{Data Availability}

\begin{table}
\begin{spacing}{1.0}
\begin{center}
 \begin{tabular}{| l | l | l |}
\hline \hline
Temp. & Homogeneous & VLE Data \\
& Density& \\
\hline \hline
263.15K &  & \cite{Fredenslund70} $^{ \dagger }$ \\
273.15K &  & \cite{Fredenslund70} $^{ \dagger }$, {\bf \cite{Muirbrook64,Muirbrook65}}, \cite{Zenner63} $^{ \dagger }$ \\
283.15K &  & {\bf \cite{Fredenslund70} }$^{ \dagger }$ \\
302.22K & \cite{Mantovani12} &  \\
\hline
\multicolumn{3}{| c |}{$^{ \dagger }$ these VLE data do not contain volumes} \\
\hline \hline
\end{tabular}
\caption{A summary of available data for \CO--O$_2$ binary mixtures. Experimental data used in
  our parameter fitting are in boldface.  \label{table:CO2O2DataSummary}}
\end{center}
\end{spacing}
\end{table}

  The availability of data for \CO --O$_2$ is
  more limited than for \CO --N$_2$. A tabulated summary of the
  relevant available data is given in Table
  \ref{table:CO2O2DataSummary}. There is a single temperature at
  which the full array of VLE information was presented, at 273.15K
  \cite{Muirbrook64,Muirbrook65}. 
The highest temperature at which any
  VLE data is presented is 283.15K, and there is no temperature at
  which both VLE and density data were given. 
To enable fitting to a second temperature we developed a volume
estimation technique for the co-existing
volumes, as detailed in \ref{section:CoexVolEst}, to fill-in where
experimental measurements were unavailable.
A summary of the data used for fitting is as
follows. For 273.15K we used VLE data by Muirbrook {\em et al.}
\cite{Muirbrook64,Muirbrook65}. For 283.15K we
used VLE data from Fredenslund {\em et al.}\cite{Fredenslund70}, with
estimated coexisting volumes.  At both temperatures we
also used pressure-volume data for pure O$_2$ from the NIST website \cite{NIST}.

\subsection{Fitting the \CO--O$_2$ Binary Model Parameters}
To fit the \CO--O$_2$ parameters, we proceeded in
 exactly the same way as for fitting the impurity parameters in the
 previous section.  We used the error function in
 eqn~\eqref{equation:ErrFnMixCO2N2} and fitted the seven impurity
 parameters  via simulated
 annealing, treating each temperature separately. We used $W=0.1$, to balance fitting between the VLE and
 homogenous phase data. 
 To obtain an expression for the temperature dependence of each
 parameter we converted our parameter values into a linear dependence
 on temperature (eqn~\eqref{equation:LinearFit}). The coefficients of
 this linear expression for each of the parameters are listed in table~\ref{table:CO2O2ParameterValues}.

\begin{table}
\begin{center}
 \begin{tabular}{| c || c | c |}
\hline 
$\theta$ &  $\alpha_0$& $\alpha_1$ \\
\hline
$a_{\mathrm{O_2}}$ &-9.3985 &10.4231 \\
$b_{\mathrm{O_2}}$ &4.62523 &-5.15272\\
$c_{\mathrm{O_2}}$ & -1.50723 &1.9847	\\
$d_{\mathrm{O_2}}$ & 12.838 &-14.292	   \\
$e_{\mathrm{O_2}}$ & 5.99309   &-6.18695 \\
$f_{\mathrm{O_2}}$ & 9.29129 &  -9.9675 \\
$g_{\mathrm{O_2}}$ &-7.61445   &8.30386  \\
\hline
\end{tabular}
\caption{Coefficients describing the temperature dependence of the
  O$_2$ model parameters in eqn~\eqref{equation:LinearFit}. \label{table:CO2O2ParameterValues}}
\end{center}
\end{table}

\subsection{Performance of the Full Model for Binary Mixtures of
 \CO and O$_2$}

\begin{figure}
\begin{center}
\includegraphics[scale=0.4]{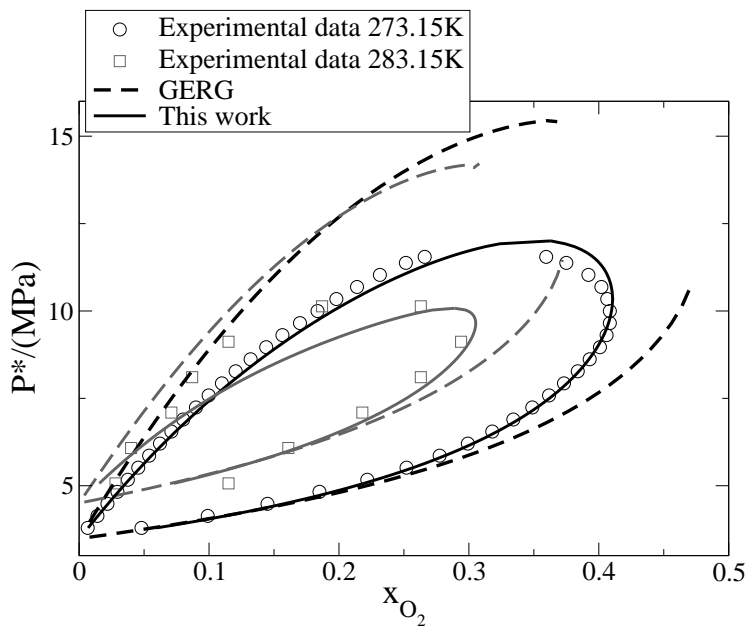}
\caption{ Comparison of experiments and model predictions for the VLE
  coexisting mole fraction for \CO-O$_2$ mixtures. Experimental data at 273.15K from
  Muirbrook {\em et al.}\cite{Muirbrook64,Muirbrook65}; and at 283.15K from Fredenslund {\em
    et al.}\cite{Fredenslund70}.  Also shown are the results from the GERG EoS
\cite{Gerg2004}.\label{figure:CO2O2CoexxP}}
\end{center}
\end{figure}

\begin{figure}[htb]
\begin{center}
\includegraphics[scale=0.4]{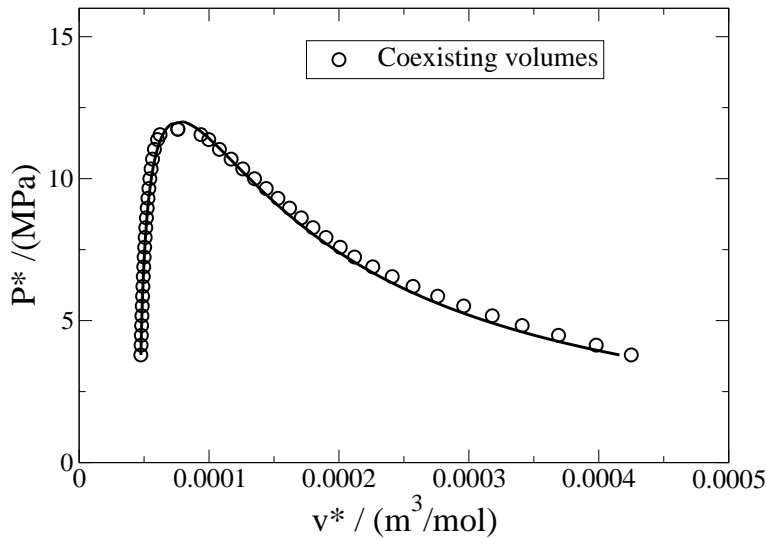}
\caption{Comparison of experiments and our model for the
  coexisting volumes for \CO-O$_2$ mixtures at 273.15K. Experimental data from
  Muirbrook {\em et al.}\cite{Muirbrook64,Muirbrook65}. \label{figure:CO2O2CoexvP}}
\end{center}
\end{figure}

Similarly, to \CO-N$_2$ mixtures we computed the VLE correlations of the
model and compared to the data used for fitting. A comparison between the experiments and our model for the coexisting
mol fraction at both temperatures is
shown in figure~\ref{figure:CO2O2CoexxP}.  Also shown are the results from the GERG
EoS. At 273.15K our model captures well the data, with a very slight
discrepancy very close to the critical point. In contrast, the GERG
dramatically over-predicts the critical point pressure and has noticeable
disagreement with the liquid mol fraction that persists over most of the
pressure range. There is a similar picture for the GERG at 283.15K,
whereas our model under-predicts somewhat the critical point at this
temperature. Overall our model produces consistently better agreement
than the GERG. We note the SAFT and PC-SAFT EoS have been compared to
these data at 288.15K\cite{Diamantonis:2013p4206} and both over-predict the
critical point pressure to a simular extent to the GERG. Finally the
EOS-CG model has also been compared with these
experiments\cite{GenertThesis2013}. The EoS-CG over-predicts the
critical point pressure by $\sim20\%$ (much less than the GERG and
SAFT models) and has better agreement to the liquid mol fraction
experiments than our model. Overall our model has superior agreement
than the GERG and the SAFT models and comprable agreement to the
EOS-CG model.

Figure~\ref{figure:CO2O2CoexvP} compares our model with measurements
of the   coexisting volumes for \CO-O$_2$ mixtures at
273.15K. Agreement is very close throughout.

\section{Carbon Dioxide--Hydrogen Binary Mixture\label{section:Fitting the CO$_2$-H$_2$ Model Parameters}}

\begin{table}
\begin{center}
 \begin{tabular}{| c || c | c |}
\hline 
$\theta$ &  $\alpha_0$& $\alpha_1$ \\
\hline
$a_{\mathrm{H_2}}$ & -0.919008 &0.910607 \\
$b_{\mathrm{H_2}}$ & -1.00994& 0.814616\\
$c_{\mathrm{H_2}}$ & -16.5057&18.3776	\\
$d_{\mathrm{H_2}}$ &0.0204381 &	   -0.0227559 \\
$e_{\mathrm{H_2}}$ &  -16.2801 &18.1264 \\
$f_{\mathrm{H_2}}$ &10.1075 & -10.2491 \\
$g_{\mathrm{H_2}}$ &  -4.5342 & 4.38617 \\
\hline
\end{tabular}
\caption{Coefficients describing the temperature dependence of the
  H$_2$ model parameters in eqn~\eqref{equation:LinearFit}. \label{table:CO2H2ParameterValues}}
\end{center}
\end{table}

Coexisting mol fractions for \CO-H$_2$ mixtures under conditions
relevant to CCS have recently been measured by  Fandi{\~n}o {\em et al.}\cite{Fandino:2015p4210}. We
fitted our model to data at 273.15K and 295.65K, to span the
temperature range of CCS pipeline operation. As the Fandi{\~n}o {\em
  et al.} data do not have corresponding coexisting volumes, we
required estimated volumes. For 273.15K we were able to validate our
volume estimates against the limited \CO-H$_2$ data from
\cite{Freitag86} at this temperature, enabling us to successfully
obtain H$_2$ parameters. However, the temperature change to 295.65K was
sufficiently large that we were unable to successfully fit the Fandi{\~n}o {\em et
  al.} mol fraction data at this temperature using our volume estimate. Thus we
allowed simulated annealing to adjust the parameters of our volume
estimation technique in situe, during the fitting, to compensate for
the lack of volume data. See \ref{sec:estimating-co-h_2} for details. 
We followed the same fitting procedure as the previous two cases and obtained the temperature dependence of each
 parameter via eqn~\eqref{equation:LinearFit}. The coefficients of
 this linear temperature rule for each of the parameters are listed in table~\ref{table:CO2H2ParameterValues}.

\begin{figure}[htb]
\begin{center}
\includegraphics[scale=0.4]{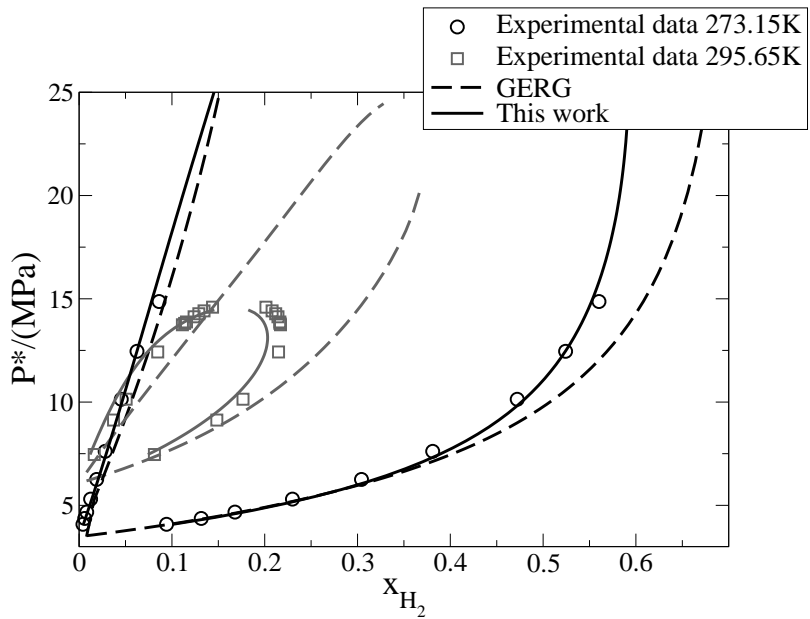}
\caption{ Comparison of experiments and model predictions for the VLE
  coexisting mole fraction for \CO-H$_2$ mixtures. Experimental data
  from Fandi{\~n}o {\em et al.}\cite{Fandino:2015p4210}. Also shown are the results from the GERG EoS
\cite{Gerg2004}.\label{figure:CO2H2CoexxP}}
\end{center}
\end{figure}

 A comparison between the experiments and our model for the coexisting
mol fraction at both temperatures is
shown in figure~\ref{figure:CO2H2CoexxP}.  Also shown are the results from the GERG
EoS. Our model captures these data very accurately, with the only
disagreement being a slight
discrepancy about the critical point for the higher temperature
data. In contrast, the GERG EoS substantially over-predicts the critical
point pressure at 295.65K, leading to large
discrepancies that persist throughout the pressure range. Furthermore, at 273.15K the GERG
EoS fails to capture the vapour mol fraction at moderate and higher pressures. Our model produces
significantly closer agreement to these data than the GERG EoS. To our
knowledge, the SAFT, PC-SAFT and EOS-CG have not been compared to
these or comparable data.

\section{Discussion}

%Summary
In this article we have developed a general framework for producing
pressure-explicit EoS for impure CO$_2$, aimed at modelling for CCS
transport. Under our approach the mixture fugacity integral, required
for coexistence calculations, can be computed from the pure fluid
fugacity and mixing rules without further integration. This assists
ongoing development of EoS in response to new data and computational
requirements, as it allows convenient replacement and modification of
terms in the EoS and mixing rules.

We used our method to generalise a recent EoS for pure \CO to binary
mixtures. We used simulated annealing to fit
model parameters to experimental data for mixtures with N$_2$, O$_2$ and H$_2$. We captured the coexistence data by
imposing a term in the fitting that penalises
differences in the predicted fugacity at the experimentally determined
coexistence points. Where volumetric measurements were unavailable for
coexistence, we developed, tested and exploited a volume estimation
method. These estimated volumes allowed us to impose appropriate
coexistence behaviour on our model.

For \CO-N$_2$ mixtures, there was comprehensive and high quality
literature data across virtually the entire CCS-pipeline window. Such data allowed us to achieve very good
 fitting with our model to both coexistence and pressure-volume data. For this system our model has comparable
 agreement to the GERG, but is slightly less accurate. However, for
 \CO-O$_2$ mixtures our model outperforms the GERG EoS, having more
 accurate coexisting mol fractions predictions at 273.15K and
 288.15K. Our model is also more accurate than the SAFT EoS at 288.15K
 and has a similar level of agreement to the recent EOS-CG\cite{GenertThesis2013}.
We also successfully compared to coexisting mol fraction data for
\CO-H$_2$ mixtures. Here our model is significantly more accurate than
the GERG EoS, particularly at higher temperatures.

%Computational complexity
A key issue for CCS modelling is the computational complexity of
EoS. Some numerical codes use lookup tables of thermophysical
properties, whereas other codes use direct, in-situe evaluation of EoS
and so require very fast EoS. Our EoS has a moderate number of terms,
with inexpensive linear relationships for the mixing rules  and the
temperature-dependence of the
impurity parameters. Furthermore,  the pressure expression of our EoS contains only rational functions and
so evaluation of this part of the EoS is computationally cheap. The
required computational efficiency and whether or not to use lookup
tables will depend upon the individual computational task. Ideally, a
new EoS would be tailored to the computational and accuracy
requirements of the application, a task towards which this work
contributes.

%Weakness of data comparison
A persistent weakness for many EoS is modelling coexistence at
pressures and temperatures in the
vicinity of the critical point. Although our model is more accurate
here than the GERG EoS for \CO-O$_2$ and \CO-H$_2$, it has small
discrepancies in this region for all impurities studied herein.
This consistent behaviour suggests a
 systematic difficulty in our current approach, either in locating
 suitable parameters to describe the vicinity of the critical pressure
 or in the mathematical terms used in the EoS; and perhaps a
 difficulty in these factors across all current EoS. We discuss below,
 improved fitting algorithms that we are currently working on to address this problem.
A further possibility to improve these issues is new mixing rules or
modified terms in the EoS. We note that our EoS framework will make
formulating these modifications straightforward and
convenient. However, we leave exploration of these ideas to
forthcoming studies.

%Weakness of our current approach
There are a number of weaknesses to our current approach. 
Firstly, there is a limit on the number of parameters that can be fitted by simulated
annealing to typical experimental data. To control the number of parameters in each
search, we fitted individual temperatures separately. Consequently, we could only
fit temperatures where VLE mol fractions and volumes were
available. Although this was mitigated somewhat by our volume
estimation techniques, we still had to exclude some
relevant data because the literature experiments were not sufficiently
comprehensive at
the relevant temperature. However, improved volume estimation or a
modified approach to fitting coexistence data may improve the
fitting, especially for impurities where mixture data are sparse.

Our approach opens up numerous possibilities to address the above weaknesses. A key issue is the effectiveness of the parameter
fitting algorithms. Effective non-linear optimisation is notoriously
difficult as parameter-search algorithms often get trapped in deep local minima
or struggle to explore effectively a complicated error
function (such as eqn~\eqref{equation:ErrFnMixCO2N2}). In this
work these factors limited the number of parameters we could fit simultaneously.
These issues could be addressed by improved fitting algorithms. For
example parallel tempering enables search algorithms to escape deep
local minima\cite{Earl:2005p4179} and Riemann manifold, Langevin and
Hamiltonian Monte Carlo methods are designed to cope with
complicated error surfaces\cite{Girolami:2011p4192}. Such improvements
are likely to allow more model parameters to be fitted
simultaneously. This will enable the temperature rules for model parameters to be
fitted directly, thus meaning that experiments spread across many
different temperature can be captured simultaneously. This may improve
the EoS's temperature-dependence around the critical point. 

There is also a central role for methods to address missing data or sparse
volumetric data. We relied on our volume estimation techniques to
stand-in for missing VLE data. A more accurate and physically-based estimate for these
volumes could be obtained from molecular simulation\cite{Deublein:2011p3873,Tenorio2015}, provided suitable
molecular force fields describing the interaction of \CO with relevant
impurities can be determined. It is also possible to treat missing VLE data as model parameters and
learn these during the fitting process, bypassing the need to
explicitly estimate missing coexistence volumes. This introduces many
new parameters to the fitting and so is contingent on solving the
problem of fitting many parameters simultaneously, either by the
techniques mentioned above, or otherwise.

Reasonable solutions to the above issues will enable extensions
of this work, targeting features that are useful to CCS
applications. These include modelling a wider range of species of
impurities, and ternary and higher order mixtures. Another future extension is
uncertainty quantifications, which will address uncertainty due to
sparsity of data for many impurities and guide the design of future experiments.

\section{Conclusions}

%Summary
In this article we have developed a general framework for producing
pressure-explicit EoS for impure CO$_2$, aimed at modelling for CCS
transport. This approach allows convenient modification of terms in
the EoS and the mixing rules. We used our method to generalise a recent EoS for pure \CO to binary
mixtures with N$_2$, O$_2$ and H$_2$, introducing 14 new model
parameters per impurity. Our model pertains to pressures up to 16MPa and
temperatures between 273K and the critical temperature of pure
CO$_2$, $T_c$.  
For \CO-N$_2$ mixtures, our model has comparable
 agreement to the GERG, but is slightly less accurate. For
 \CO-O$_2$ mixtures our model outperforms the GERG EoS, for
 coexistence data  at 273.15K and
 288.15K. Here, our model is also more accurate than the SAFT EoS at 288.15K
 and has a similar level of agreement to the recent EOS-CG.
We also successfully compared to coexisting mol fraction data for
\CO-H$_2$ mixtures. Here our model is significantly more accurate than
the GERG EoS, particularly at higher temperatures.
We note that it is possible that some pipelines may operate
above $T_c$, in some regions, which our pure \CO model does not
currently account for. Fitting this region is more straightforward
than the subcritical region due to the lack of coexistence. 
Furthermore, pipeline failure modelling requires an EoS that is
accurate down to the \CO triple-point temperature, which is below the
range we have explored here. Although our methodology is appropriate
to these extensions, we leave them to future work.

\section*{Acknowledgments}
This work was supported by funding from RCUK and RWE nPower. RSG was supported by the Materials for Next Generation \CO Transport Systems (MATTRAN) project, funded by the EPSRC (EP/G061955/1). The authors thank the MATTRAN project partners and Prof Roland Span for very useful discussions concerning this work.

\appendix
\section{ Mixture Fugacity in Equations of State}\label{sec:mixt-fugac-equat}
We begin with the following fugacity expressions from Orby {\em et al}\cite{Orbey98}: equation (2.3.9) for the fugacity coefficient of a pure fluid, $\ln\phi$, is
\begin{equation}
  \label{eq:3a}
\ln\left[\frac{f^*_i(T^*,V^*)}{P^*}\right]=\ln\phi=\frac{1}{RT^*}\int_{\infty}^{V^*}\frac{RT^*}{V'^*
  }-\frac{P^*}{N}\ud V'^* -\ln Z+(Z-1),
\end{equation}
where $V^*$, $N$ and $Z=P^*v^*/RT^*$ denote the volume, total number of moles
and compressibility, respectively; and equation (2.3.1) for the fugacity coefficient of species $i$ in a mixture is
\begin{equation}
  \label{eq:1}
  \ln\left[\frac{f^*_i}{P^*x_i}\right]=  \ln\bar{\phi}_i=\frac{1}{RT^*}\int_{\infty}^{V^*}\frac{RT^*}{V'^*}-\left(\dpart{P^*}{N_i}\right)_{T^*,V^*,N_{j\neq
      i}}\ud V'^*-\ln Z,
\end{equation}
where $N_i$ and $x_i$ are the total number of mols and mol fraction, respectively, of species $i$. All
variables in dimensional units are notated with stars.

Pressure-explicit EoS are of the form $P^*=P(v^*,\vect{\theta}^*(N_i))$, where $v^*=\frac{V^*}{N}$, so we can use the chain rule to write
  \begin{equation}
    \label{eq:2}
\left(\dpart{P^*}{N_i}\right)_{T^*,V^*,N_{j\neq i}}= -\dpart{P^*}{V^*}\frac{V^*}{N_T}+\sum_{j=1}^{N_p}\dpart{P^*}{\theta^*_j}\dpart{\theta^*_j}{N_i},
  \end{equation}
where $N_p$ is the number of model parameters.
Substituting into equation~\eqref{eq:1} gives
\begin{equation}
  \label{eq:3}
  \ln\bar{\phi}_i=\left(\frac{1}{RT^*}\int_{\infty}^{V^*}\frac{RT^*}{V'^*}+\dpart{P^*}{V'^*}\frac{V'^*}{N_T}\ud V'^*-\ln Z \right)-\frac{1}{RT^*}\sum_{j=1}^{N_p}\dpart{\theta^*_j}{N_i}\int_{\infty}^{V^*} \dpart{P^*}{\theta^*_j} \ud V'^*.
\end{equation}
Integration by parts shows that the term inside the brackets is equal to $\ln\phi$ (by eqn~\eqref{eq:3a}). Furthermore, 
defining $F$ to be the integral from eqn~\eqref{eq:3a}, i.e. $F=\frac{1}{RT^*}\int_{\infty}^{V^*}\frac{RT^*}{V'^*}-\frac{P^*}{N}\ud V'=\ln\phi-((Z-1)-\ln Z)$, allows us to write.
\begin{equation}
  \label{eq:5}
  -N\dpart{F}{\theta^*_j}=\frac{1}{RT^*}\int_{\infty}^{V^*}\dpart{P^*}{\theta^*_j}\ud V'^*.
\end{equation}
Substituting \eqref{eq:5} into equation~\eqref{eq:3} gives
\begin{equation}
  \label{eq:8}
  \ln\bar{\phi}_i=\ln\phi+N\sum_{j=1}^{N_p}\dpart{F}{\theta^*_j}\dpart{\theta^*_j}{N_i}.
\end{equation}

The mixing rule will be of the form $\vect{\theta}^*=\vect{\theta}^*(x_i,\{x_{k\neq i}\})$, where $x_i$ is the mol fraction of species $i$ and the $x_k$ are the mol fractions of all of the remaining species. Hence, we must now convert the derivative $\dpart{\theta^*_j}{N_i}$ in eqn~\eqref{eq:8} into a derivative with respect to the mol fractions. We have $x_i=\frac{N_i}{N_i+N_{other}}$ where $N_{other}$ is the total number of mols of all species other than $i$. Similarly for $k\neq i$, we have $x_k=\frac{N_k}{N_i+N_{other}}$.
Differentiating $\theta^*_j$ with respect to $N_i$ gives
\begin{equation}
  \label{eq:9}
      \dpart{\theta^*_j}{N_i}=\dpart{\theta^*_j}{x_i}\dpart{x_i}{N_i}+\sum_{k\neq i}\dpart{\theta^*_j}{x_k}\dpart{x_k}{N_i}.
\end{equation}
Differentiating the expressions for $x_i$ and $x_k$ allows us to write
\begin{equation}
  \label{eq:11}
      N \dpart{x_i}{N_i}=1-x_i\;\; \mbox{ and }\;\; N\dpart{x_k}{N_i}=-x_k \;\;(\mbox{for } k\neq i).
\end{equation}
Substituting equations~\eqref{eq:9} and \eqref{eq:11} into equation~\eqref{eq:8} gives
\begin{equation}
  \label{eq:13}
  \ln\bar{\phi}_i=\ln\phi+\sum_{j=1}^{Np}\dpart{F}{\theta^*_j}\left(\dpart{\theta^*_j}{x_i}-\sum_{k=1}^{N_{sp}}x_k\dpart{\theta^*_j}{x_k}\right).
\end{equation}

We note here that, since the fugacity coefficients are dimensionless, then the mixing rules $\vect{\theta}^*$ are the only dimensional quantities in eqn~\eqref{eq:13}. Furthermore, because of the arrangement of the $\vect{\theta}^*$ terms in \eqref{eq:13}, the expression is unchanged by any constant scaling of the parameters $\vect{\theta}^*$. Thus we conclude that the form of eqn~\eqref{eq:13} is unchanged by any non-dimensionalisation that applies a constant scaling to the model parameters.

\section{Estimation Molar Volumes for Mixtures}\label{section:CoexVolEst}

 As noted previously, we observed that a lot of literature VLE data
 lacks measurement of the coexisting volumes. 
%Such data points are noted in the Appendices under the ``Vol. Est.''
%column, by which we meant that even though the data point did not
%contain the volumetric information, it might be still be usable if we
%could somehow introduce a suitable method of volume estimation. 
We thus developed from scratch a novel process to estimate both the
liquid and vapour volumes in compensation for those data missing in
the literature. This was primarily because our method of fitting the
parameters required all elements of
the thermodynamic description (temperature, pressure, volumes and
compositions) to be present. We acknowledge that experimental
measurements are clearly preferable to estimated
volumes and have only used this technique when measurements are not available.

\subsection{Estimating the Coexisting Liquid Volume}

 We estimated the mixture liquid volume $\tilde{u}_{\mathrm{liq,MIX}}$
 by taking a weighted average of the \CO volume at the pressure of interest
 and the pure impurity volume at the same pressure, weighted by the
 impurity concentration raised to an empirical power $\alpha$. The
 pure data was available in all cases using standard reference
 EoS\cite{Span96,Span:2000p4190,Leachman:2009p4189,Schmidt:1985p4191}
 accessed from the NIST website\cite{NIST}. Our empirical expression
 for the estimated volume of the coexisting
mixture, $\tilde{v}_{\mathrm{liq,MIX}}$, at the required temperature and pressure is,
\begin{equation}\label{equation:vBPEst}
\tilde{v}_{\mathrm{liq,MIX}} = (1 - x_{\mathrm{liq}}^\alpha) v_{\mathrm{CO_2}} + x_{\mathrm{liq}}^\alpha v_{\mathrm{imp}},
\end{equation}
where $ x_{\mathrm{liq}}$
is the impurity concentration of the coexisting liquid,
$v_{\mathrm{CO_2}}$ and $v_{\mathrm{imp}}$ are the pure \CO and pure impurity molar volume at the
temperature and pressure of interest, and $\alpha$ is  an
empirically determined constant, included to calibrate this estimation
for different mixtures, whose values for different mixtures are given in Table \ref{table:BubDewConst}.

\subsection{Estimating the Coexisting Vapour Volume}
\begin{figure}[htb]
\begin{center}
\includegraphics[scale=0.5]{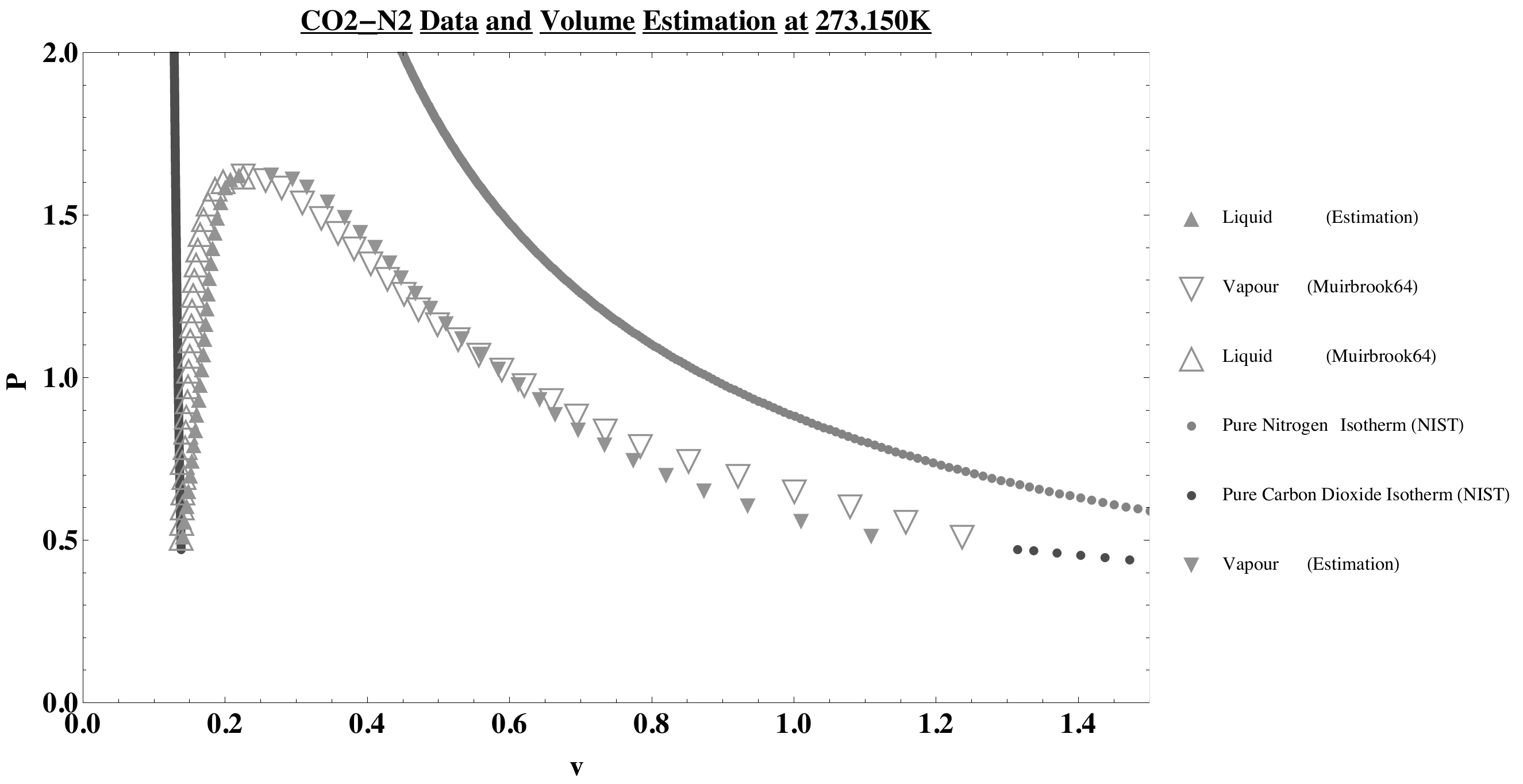} \includegraphics[scale=0.5]{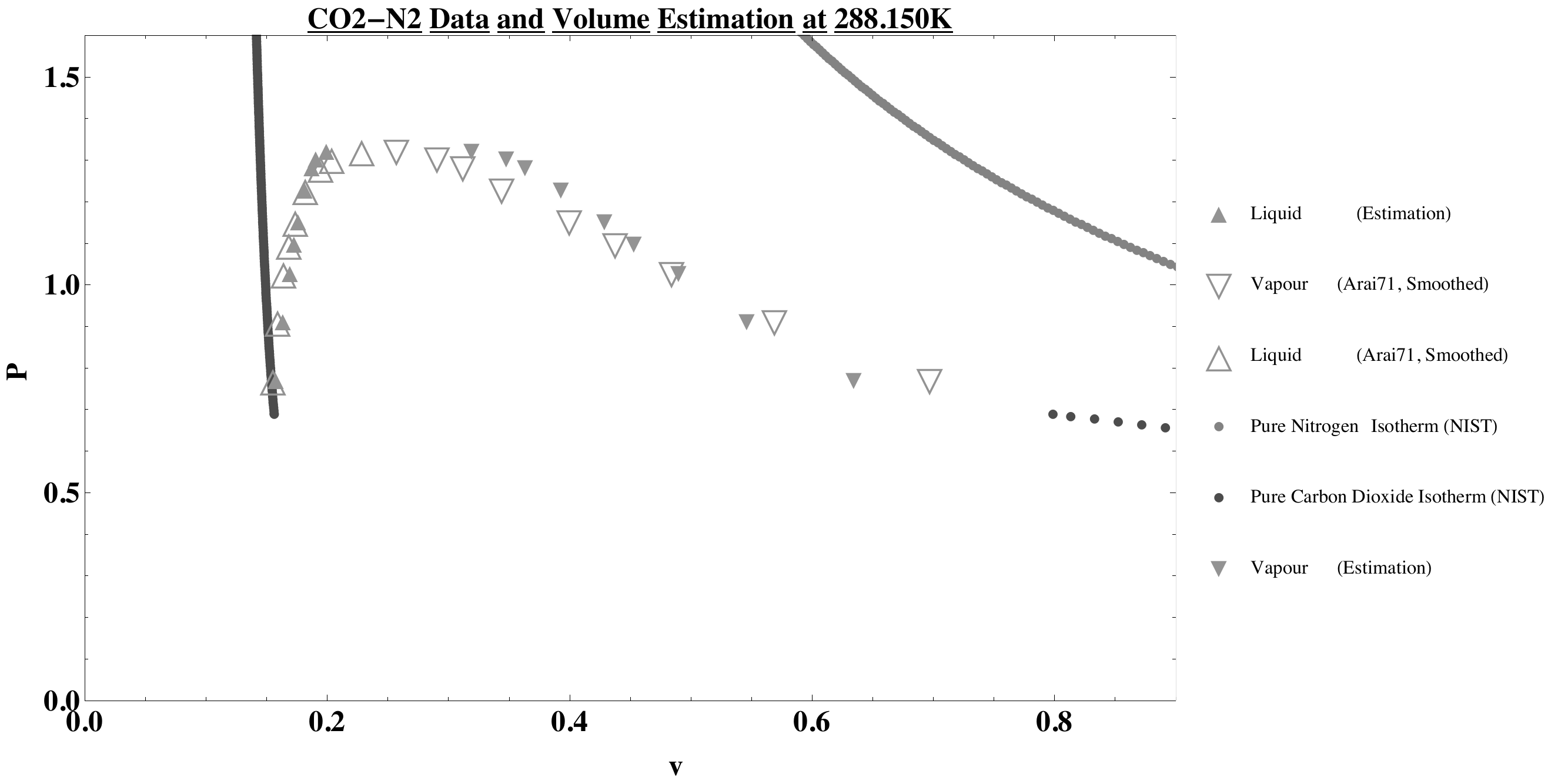}
\caption{Behaviour of the volume estimation method compared to
  literature data at 273.15K (top) \cite{Muirbrook64} and 288.15K (bottom) \cite{Arai71} for a \CO--N$_2$ mixture \label{figure:VolEstCO2N2}}
\end{center}
\end{figure}

\begin{figure}[htb]
\begin{center}
\includegraphics[scale=0.5]{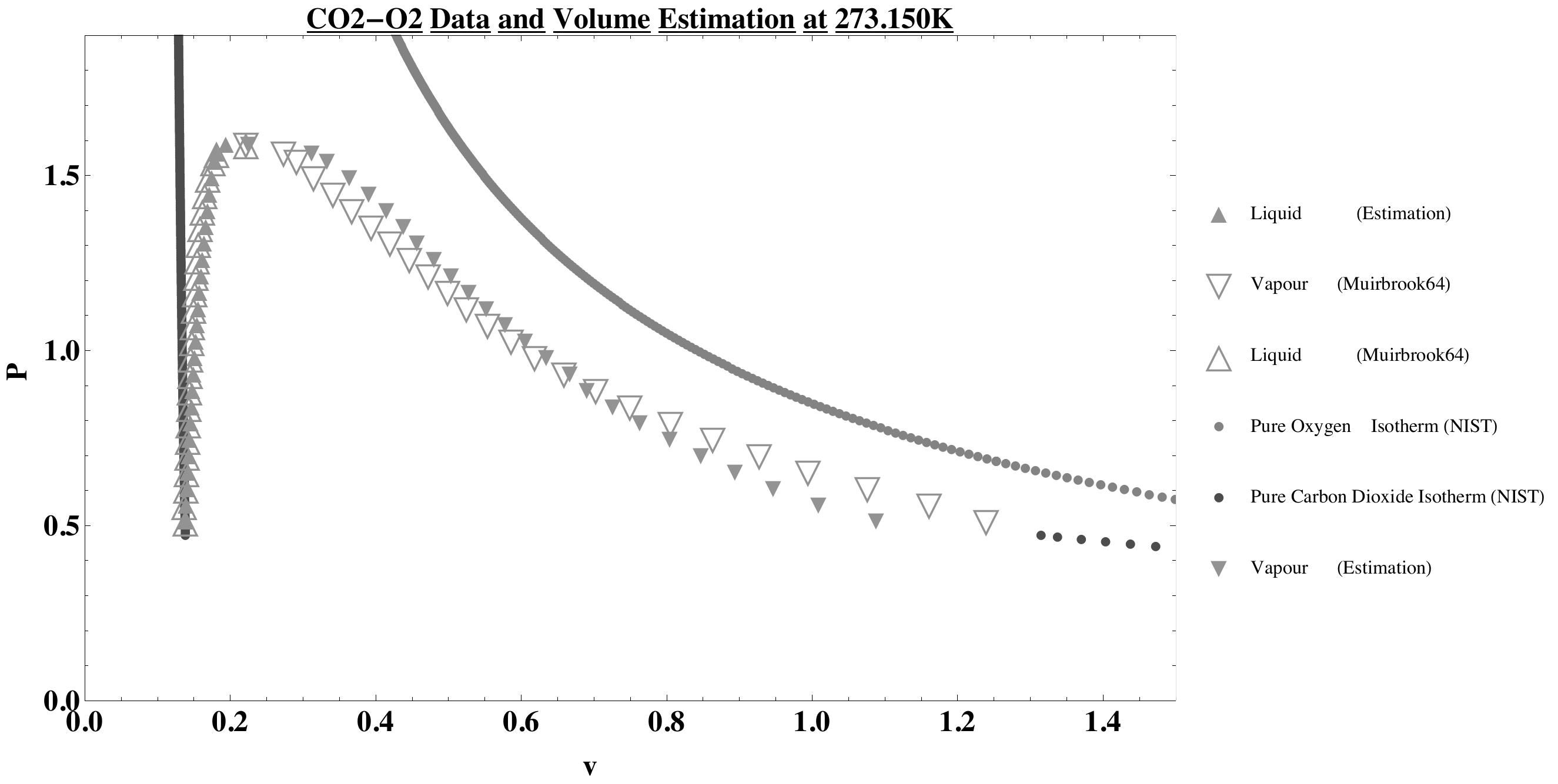}
\caption{Behaviour of the volume estimation method compared to literature data at 273.15K for a \CO--O$_2$ mixture \label{figure:VolEstCO2O2T273p150}}
 \end{center}
\end{figure}

\begin{figure}[htb]
\begin{center}
\includegraphics[scale=0.5]{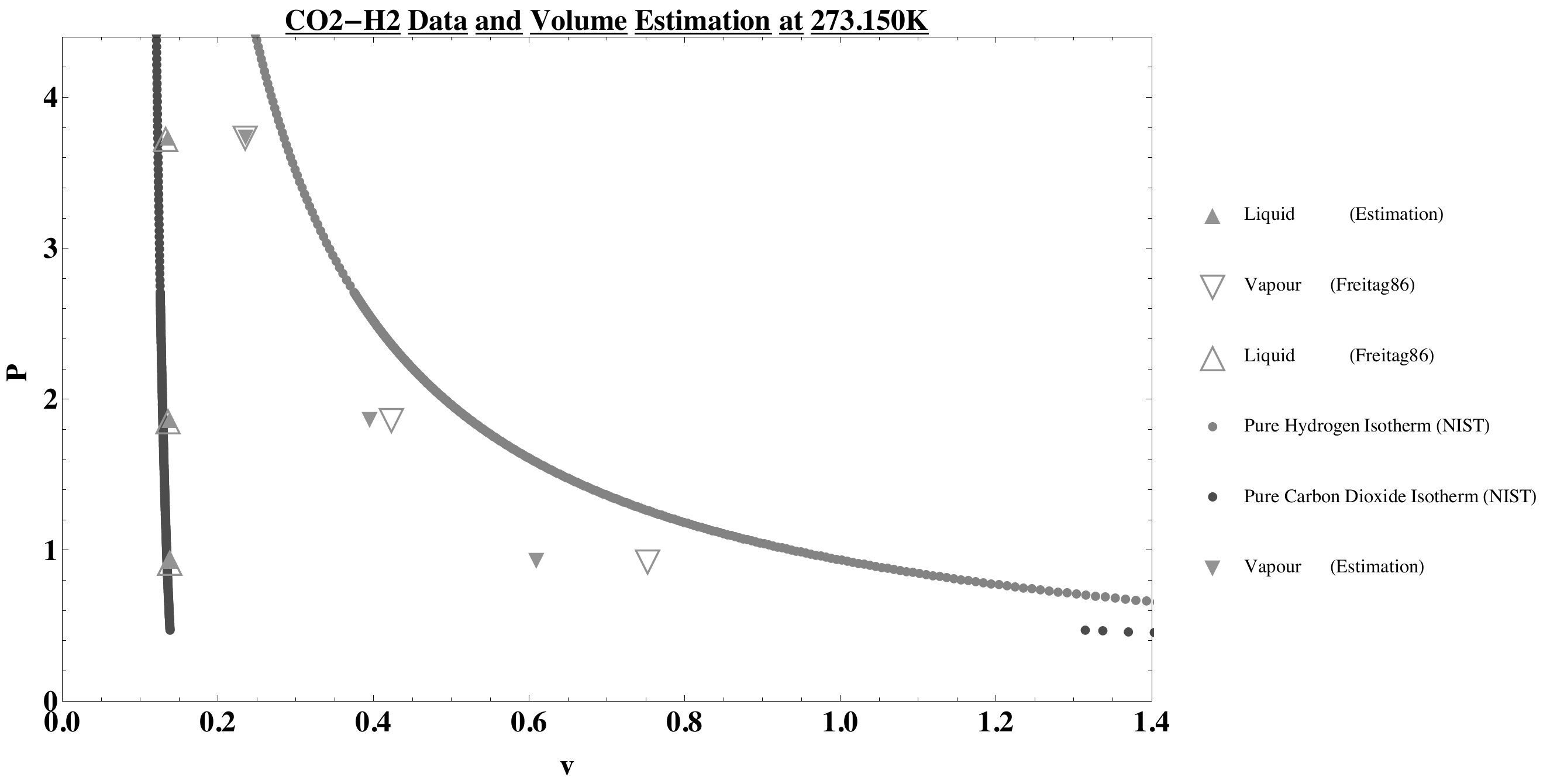}
\caption{Behaviour of the volume estimation method compared to literature data at 273.15K for a \CO--H$_2$ mixture \label{figure:VolEstCO2H2T273p150}}
\end{center}
\end{figure}

The simple weighted average of eqn~\eqref{equation:vBPEst} is
ineffective at estimating the coexisting vapour volume, due to the
higher impurity fraction and its influence on the \CO in the vapour
phase. Thus a slightly more complicated expression is required. We
used the following expression to estimate the coexisting vapour volume
for the vapour,$\tilde{v}_{\mathrm{vap,MIX}}$
\begin{equation}\label{equation:vDPEst}
\tilde{v}_{\mathrm{vap,MIX}} = x_{\mathrm{vap}}^{\beta} v_{\mathrm{CO_2}} + v_{\mathrm{imp}} \left(\frac{ v_{\mathrm{vap,CO_2}} -v_{\mathrm{liq,CO_2}}}{v_{\mathrm{imp}}
  - v_{\mathrm{liq,CO_2}}} (1 - x_{\mathrm{vap}}^{\beta}) + (x_{\mathrm{vap}} - x_{\mathrm{liq}})^{\gamma}\right),
\end{equation}
where $ x_{\mathrm{vap}}$ is the impurity concentration of the
coexisting vapour, $v_{\mathrm{vap,CO_2}}$ and $v_{\mathrm{liq,CO_2}}$
  are the coexisting vapour and liquid volumes for pure \CO 
,and $\beta$ and $\gamma$ are two further
empirically determined constants, given in Table
\ref{table:BubDewConst}. The three empirical exponents depend only on
the type of impurity and we determined these by fitting to the
coexistence volume measurements used in this work.
\begin{table}[htbp]
\begin{spacing}{1.0}
\begin{center}
 \begin{tabular}{| c || c | c | c |}
\hline \hline
Mixture & $\alpha$ & $\beta$ & $\gamma$ \\
\hline \hline
CO$_2$--N$_2$ 			& 1.25	& 0.68 & 1.01	\\
%CO$_2$--H$_2$ 			& 1.51	& 0.48 & 1.19	\\
CO$_2$--O$_2$ 			& 1.55	& 0.47 & 0.73	\\
CO$_2$--H$_2$ 			& 1.51	& 0.48 & 1.19	\\
\hline \hline
\end{tabular}
\caption{A summary of the values for the empirical exponents
  obtain by fitting to coexisting volume measurements.\label{table:BubDewConst}.}
\end{center}
\end{spacing}
\end{table}
When fitting our EoS, we used experimental measurements for the
coexisting volumes whenever available and used estimated volumes
otherwise. The benefit of this method is that it
allows a substitute value for the missing coexisting volumes, which
often were not quoted, to be estimated. The technique requires only the pure density
data, which was readily available through the NIST website\cite{NIST} and the
coexisting molar fractions, which are widely reported from
experiments. A comparison between our estimation technique and
co-existing volume measurements used in this work are shown in
figures~\ref{figure:VolEstCO2N2}, \ref{figure:VolEstCO2O2T273p150} and \ref{figure:VolEstCO2H2T273p150}.

\subsection{Estimating the \CO-H$_2$ volumes close to the critical
  point}
\label{sec:estimating-co-h_2}
For 273.15K we were able to validate our
volume estimates against the limited \CO-H$_2$ data from
ref \cite{Freitag86} at this temperature, enabling us to successfully
obtain H$_2$ parameters (see figure~\ref{figure:VolEstCO2H2T273p150}). However, the temperature step to 295.65K was
sufficiently large that we were unable to successfully fit the Fandi{\~n}o {\em et
  al.} VLE data at this temperature using our volume estimate. Thus we
took a slightly more flexible form for the volume estimate
\begin{equation}
  \label{eq:7}
  \tilde{v}_{\mathrm{vap,MIX}}(P) =(v_{\mathrm{vap,CO_2}}-v_{\mathrm{crit}})\left(\frac{P_{\mathrm{crit}}-P}{P_{\mathrm{crit}}-P_{\mathrm{CO_2}}}\right)^{\phi}+v_{\mathrm{crit}},
\end{equation}
and
\begin{equation}
  \label{eq:7b}
  \tilde{v}_{\mathrm{liq,MIX}} (P)=(v_{\mathrm{liq,CO_2}}-v_{\mathrm{crit}})\left(\frac{P_{\mathrm{crit}}-P}{P_{\mathrm{crit}}-P_{\mathrm{CO_2}}}\right)^{\phi\left(\frac{v_{\mathrm{crit}} - v_{\mathrm{liq,CO_2}}}{v_{\mathrm{vap,CO_2}}- v_{\mathrm{crit}}}\right)}+v_{\mathrm{crit}},
\end{equation}
where $P_{\mathrm{CO_2}}$ is the pure \CO coexistence pressure at
295.65K and $P_{\mathrm{crit}}$ is the critical point pressure for
\CO-H$_2$ mixtures at 295.65K, which  Fandi{\~n}o {\em et
  al.} evaluated as $P_{\mathrm{crit}}^*=14.655$ MPa from their data. The parameters to be
fitted are the exponent $\phi$ and the mixture critical volume, $v_{\mathrm{crit}}$.
We
allowed simulated annealing to adjust the parameters of these two
parameters, at the same time as the model parameters, during the
fitting at 295.65K. This process gave values of $v_{\mathrm{crit}}=0.320291$ and $\phi=1.02222$.

%\bibliography{ThesisReferences}
%\bibliographystyle{/Users/pmzrsg/Dropbox/Bibtex.dir/nature}

\end{document}